\documentclass[4pt, a4paper,prd,twoside,superscriptaddress, nofootinbib, twocolumn,groupedaddress]{revtex4-1}
\usepackage{graphics}
\usepackage[pdftex]{graphicx}
\usepackage{amsmath, amssymb}
\usepackage{color}
\usepackage[usenames,dvipsnames]{xcolor}

\makeatletter
\newcommand*{\rom}[1]{\expandafter\@slowromancap\romannumeral #1@}
\makeatother
\usepackage{setspace}
\usepackage{setspace}
\usepackage{amsmath}
\usepackage{latexsym}
\usepackage{amssymb}
\usepackage{graphicx,color}
\usepackage{epstopdf}
\usepackage{graphicx}
\usepackage{setspace}
\usepackage{makeidx} 
\usepackage{lmodern}
\usepackage[showframe=false]{geometry}
\usepackage{enumitem}

\usepackage[T1]{fontenc}
\usepackage[utf8]{inputenc}
\usepackage{color}
\usepackage{booktabs}
\usepackage{units}
\usepackage{amsmath}
\usepackage{amssymb}
\usepackage{esint}

\pdfpageheight\paperheight
\pdfpagewidth\paperwidth

\usepackage[unicode=true,pdfusetitle, bookmarks=true,bookmarksnumbered=false,bookmarksopen=false, breaklinks=false,pdfborder={0 0 0},backref=false,colorlinks=true]{hyperref}
\hypersetup{citecolor=blue,filecolor=blue,linkcolor=blue,urlcolor=blue}

\def\no{\nonumber}

\def\bea{\begin{eqnarray}}  
\def\eea{\end{eqnarray}}

\def\be{\begin{equation}}
\def\ee{\end{equation}}

\def\no{\nonumber}

\def\H0{H_{0}}

\def\f''{f''}

\def\q''{q''}

\def\etaf''{\eta_{f''}}

\def\etaq''{\eta_{q''}}

\def\stuck{\text{St\"uckelberg}}
\makeatletter

\newcommand{\Rmnum}[1]{\expandafter\@slowromancap\romannumeral #1@}
\makeatother

\usepackage{anysize}
\marginsize{2.5cm}{2cm}{1cm}{3.5cm}

\usepackage{natbib}
\begin{document}


\title{On the UV structure of quantum unimodular gravity}

\author{Ippocratis D. Saltas \vspace{0.2cm}}
\email{isaltas@fc.ul.pt}
\affiliation{Instituto de Astrofísica e Ciências do Espaço, Faculdade de Ciências, Campo Grande, PT1749-016 Lisboa, Portugal}
\affiliation{\vspace{0.01cm}}
\affiliation{School of Physics \& Astronomy, University of Nottingham, Nottingham, NG7 2RD, United Kingdom}


\hoffset = -1cm
\textwidth = 19cm

\begin{abstract}
It is a well known result that any formulation of unimodular gravity is classically equivalent to General Relativity (GR), however a debate exists in the literature about this equivalence at the quantum level. In this work, we investigate the UV quantum structure of a diffeomorphism invariant formulation of unimodular gravity using functional renormalisation group methods in a Wilsonian context. We show that the effective action of the unimodular theory acquires essentially the same form with that of GR in the UV, as well as that both theories share similar UV completions within the framework of the asymptotic safety scenario for quantum gravity. Furthermore, we find that in this context the unimodular theory can appear to be non--predictive due to an increasing number of relevant couplings at high energies, and explain how this unwanted feature is in the end avoided. 
\end{abstract}

\keywords{unimodular gravity, quantum gravity, renormalisation group, asymptotic safety}

\maketitle

\section{Introduction}

An old and on the same time simple attempt of attacking the cosmological constant problem has been unimodular gravity \cite{Weinberg:1988cp,vanderBij:1981ym,Buchmuller:1988wx,Buchmuller:1988yn,Henneaux:1989zc,Ng:1990rw,Ng:1990xz,Smolin:2009ti}. The fundamental idea underlying unimodular gravity is to modify the classical gravitational dynamics in a way that the cosmological constant is disentangled from the equations of motion, by making the determinant of the metric non--dynamical, or equivalently by requiring that its variation with respect to the metric field equal to zero. As we will also discuss later, the latter requirement restricts the full diffeomorphism symmetries of General Relativity (GR) to only the transverse diffeomorphism ones. 

Unimodular gravity does not succeed in eliminating the cosmological constant term from the classical dynamics; a cosmological constant term appears back into the classical equations as a constant of integration, by use of the Bianchi identities, and the classical dynamics turn out to be exactly the same with those of GR. In this context, it is essentially the conceptual problem of the cosmological constant that changes; in unimodular gravity, the cosmological constant problem accounts to tuning an integration constant to the required value, instead of a coupling that appears in the action through appropriate renormalisation conditions. For a detailed and recent discussion on the extend to which unimodular gravity can provide a ''new perspective" to the cosmological constant problem see Ref. \cite{Padilla:2014yea}.



Unimodular gravity attracted also a lot of attention in the context of quantum gravity, as a possible way of solving the problem of time in this context \cite{Smolin:2010iq,Unruh:1988in,Unruh:1989db,Bombelli:1991jj,Kuchar:1991xd}, or as a way of avoiding the conformal anomalies \cite{Alvarez:2013fs,Alvarez:2014qca}. Although it has been a well known fact since a long time ago that, at the classical level unimodular gravity is  equivalent to standard GR with a cosmological constant, the quantum--mechanical equivalence between the two theories has been a matter of debate, with contradicting results appearing in the literature. In particular, Ref. \cite{Alvarez:2005iy} studied the quadratic actions of standard GR and its unimodular counterpart around flat space-time, enforcing the constancy of the metric's determinant from the onset, and argued that in principle quantum effects can allow one to discriminate between the two theories at the experimental level, while in a similar setting Ref. \cite{Alvarez:2012px} argued that at the quadratic level of fluctuations around generic backgrounds the free energy of unimodular gravity agrees with that of GR. Furthermore, in Ref. \cite{Fiol:2008vk} it was claimed that the two theories are equivalent at the perturbative level for asymptotically flat space times, but inequivalence was found for semi-classical non--perturbative quantities. Ref. \cite{Padilla:2014yea} introduced a diffeomorphism invariant unimodular formulation of GR, through the introduction of appropriate Lagrange multiplier and $\stuck$ fields, and argued that quantum--mechanical equivalence can be established provided that only the metric field couples to external sources in the path integral. 
In a different setup, Ref. \cite{Eichhorn:2013xr} started from a version of the Einstein--Hilbert action, without a cosmological constant, which was made invariant under transverse diffeomorphism symmetry by construction and in particular, by acting upon the gauge fixing sector of the effective action with an appropriate transverse operator. Then, in the context of the Exact Renormalisation Group (ERG), the author showed that the action exhibited an attractive, non--trivial UV fixed point for Newton's coupling $G$. It is important to point out that in the latter work the cosmological constant was left out of the action, due to the enforcement of the transverse diffeomorphism symmetry.

The aim of this work will be to analyse the UV structure of quantum unimodular gravity in the context of the ERG, with the starting point being a unimodular formulation of GR which is by construction diffeomorphism invariant, as that was introduced in Ref. \cite{Padilla:2014yea}. In particular, the action will consist of the usual Einstein--Hilbert sector, as well as of a new sector constructed from five new field variables apart from the metric, namely a Lagrange multiplier and a set of four appropriate $\stuck$ fields, acting as to restore the full diffeomorphism symmetries of the theory. In this setup, the unimodularity condition will appear as an an on--shell condition implemented through the Lagrange multiplier $\lambda(x)$ and $\stuck$ fields, while it is important to point out that the quantum fluctuations we will compute will in principle be of off--shell nature. What is more, since we will be studying a fully diffeomorphism invariant theory, the cosmological constant will be also introduced in the action. As mentioned also earlier, in Ref. \cite{Padilla:2014yea} an argument was presented for the quantum equiavelence between GR and its unimodular counterpart based on decoupling the lagrange multipler and $\stuck$ fields from the sources in the generating functional; in this work, our approach will be different in this regard, since we shall treat all fields equally at the quantum level allowing them to couple to appropriate sources. 

In this context, and using as our main tool the functional RG, we will calculate the (non--perturbative) RG flow of the effective action for the unimodular theory, and proceed with studying the associated UV quantum structure, assuming the scenario of asymptotic safety. Our analysis will show that the unimodular effective action and that of standard GR acquire a similar form at sufficiently high energies in the UV. In particular, we will show that the novel interactions in the unimodular action (compared to standard GR), i.e the interactions in the St\"uckelberg sector, turn off at sufficiently high energies, while the Einstein--Hilbert sector of the action, spanned by Newton's and cosmological constant, acquires appropriate fixed-point and eigenvalues sufficiently close to those of GR. We should stress that the above results concern the regime of sufficiently high energies, suggesting that the two theories essentially share similar UV completions, and in general do not imply the quantum--mechanical equivalence between the unimodular theory and GR; in fact, and as we will also discuss later on, as one moves towards lower energies, the interactions in the $\stuck$ sector will in principle become important, leading to different quantum structure and phenomenology for the two theories.

Our analysis will further reveal that the unimodular theory appears to be non--predicitve, due to the apparent increasing  number of relevant couplings in the St\"uckelberg sector. We will discuss the root behind this issue, and how this undesired situation is finally avoided. 




We structure the paper as follows: In section \ref{sec:Classical-UM1} we review some fundamental aspects of unimodular gravity at the classical level, as well as introduce the unimodular action our quantum analysis will be based on. In section \ref{sec:part2} we introduce our main tools for the quantum analysis, and then proceed with calculating the (non--perturbative) RG flow equation for the unimodular action. After presenting some of its key properties in this context, we analyse how these compare with those of standard GR, and discuss the consequences about the equivalence between the two theories. Our conclusions are discussed in section \ref{sec:Conclusions}, while useful explicit formulas and calculations are presented in the appendix.


\section{Action setup} \label{sec:Classical-UM1}
In this section we shall first review very briefly the fundamental idea behind unimodular gravity, and we will then introduce the diffeomorphism invariant unimodular action, first introduced in Ref. \cite{Padilla:2014yea}, and which will be the starting point for our quantum analysis. 

The original motivation of unimodular gravity accounted to disentangling the cosmological constant from the gravitational equations of motion, or in other words enforce its contribution to be zero, by requiring that the metric's determinant $g \equiv \det g_{\mu \nu} $ is non--dynamical,
\be
\sqrt{-g} = \epsilon_0,  \label{UnimodCondition}
\ee
with $\epsilon_0$ a constant scalar density. The unimodularity condition (\ref{UnimodCondition}) is equivalent to requiring that the variation of the determinant $g$ with respect to the metric yields zero, disentangling this way the cosmological constant coupling from the classical equations of motion. It is well known that the enforcement of the unimodularity condition (\ref{UnimodCondition}) implies that the set of allowed diffeomorphism transformations of the theory is now restricted only to the transverse ones, i.e. the theory becomes transverse diffeomorphism invariant.



The condition (\ref{UnimodCondition}) can be implemented at the level of the action in a straightforward manner through the introduction of a Lagrange multiplier $\lambda(x)$, as was done in Refs \cite{Buchmuller:1988yn,Buchmuller:1988wx}. However, the resulting action is invariant under the restricted group of transverse diffeomorphisms, since the full symmetries will be broken through the quantity $\epsilon_0$.

In this work, the starting point for our analysis will be the unimodular action introduced in Ref. \cite{Padilla:2014yea}, where the unimodularity condition was implemented at the level of the action, but in a way that the full diffeomorphism symmetries are preserved. The action, which generalised the actions introduced in Refs \cite{Buchmuller:1988wx,Buchmuller:1988yn, Kuchar:1991xd}, introduces apart from a Lagrange multiplier $\lambda(x)$, a set of four $\stuck$ fields $\phi^{\alpha}(x) = \{\phi^{1}(x), \ldots, \phi^{4}(x)\}$, the latter acting in a way as to restore the full diffeomorphism symmetry of the theory. It reads as \cite{Padilla:2014yea}
\be
S = \int d^{4}x  \sqrt{-g} \left[ \frac{R}{16 \pi G}  +  \lambda(x)f(\psi) + q(\psi)  \right], \label{EH+lambda+phi2}
\ee
with $f(\psi)$ and $q(\psi)$ in principle arbitrary functions, and the scalar $\psi$ defined as
\be
\psi(x) \equiv \frac{ \left|J^{\alpha}{}_{\beta}\right| }{\sqrt{-g} }.
\ee
$\left| J^{\alpha}{}_{\beta}\right|$ is the determinant of the $\stuck$ Jacobian, $J^{\alpha}{}_{\beta}\equiv  \frac{\partial \phi^{\alpha}(x)}{\partial x^{\beta}}$, defined as
\begin{align}
& \left| J^{\alpha}{}_{\beta}\right| = 4! \delta^{[\alpha}_{\mu}\delta ^{ \beta}_\nu \delta^{ \gamma}_\kappa \delta^ {\delta]}_{\lambda} J^{\mu}{}_{\alpha} J^{\nu}{}_{\beta} J^{\kappa}{}_{\gamma} J^{\lambda}{}_{\delta}. \label{Jab-def}
\end{align}

The more familiar formulation of unimodular gravity corresponds to the choice $q(\psi)= 0$, and $\sqrt{-g}f(\psi) = \sqrt{-g}(1 - \psi)$, with the (dynamical) scalar $\psi$ now playing the role of $\epsilon_0$.
The classical dynamics of the unimodular action (\ref{EH+lambda+phi2}), have been discussed in Ref. \cite{Padilla:2014yea}, and as one would expect, they are exactly the same as those of standard GR with a cosmological constant. We refer to Ref. \cite{Padilla:2014yea} for more details.


\section{Quantum unimodular gravity} \label{sec:part2}
In this section we move on to study the UV quantum structure of the diffeomorphism invariant unimodular theory (\ref{EH+lambda+phi2}), as well as how it compares with that of standard GR. 
Our tool in this direction will be the functional RG in a Wilsonian context, based on an Exact RG Equation (ERGE). The power of the ERG is that it provides a unifying way to study the (non--perturbative) quantum dynamics from arbitrarily small to arbitrarily high energy/cut--off scales. The ERGE captures the essence of the Wilsonian approach to RG, where the quantum dynamics are studied by integrating out quantum fluctuations shell-by-shell in momenta. What is more, the ERGE being an exact equation, is a non--perturbative equation and no requirement about the smallness of the couplings has to be made.  

Our starting point is the path integral given by
\be 
Z[J]=\int Dg_{\mu\nu}D\phi^\alpha D\lambda \;  e^{  iS[\Phi_A]+   i\int_{}^{} J_{A} \Phi_{A}  + \Delta S_{k}   } \label{PI}
\ee
where $S[\Phi_A]$ corresponds to the action (\ref{EH+lambda+phi2}), $\Phi_A = \{ g_{\mu \nu}(x), \lambda(x), \phi^{\alpha}(x) \}$ are the field variables of the action, and $J_{A} \Phi_{A}$ implies a summation of the coupling of each of the fields $\Phi_A$ to appropriate external sources $J_A$. 
Defining $k$ to be an infrared cut--off,  the term $\Delta S_k$ plays the role of a scale-dependent infrared regulator, suppressing momenta lower than $k$, and implementing the Wilsonian idea of integrating out momenta shell by shell \cite{Gies:2006wv, Reuter:2012id, Pawlowski:2005xe, Litim:2008tt}. Denoting with $\bf{\Phi}$ any (matrix-valued) dynamical field of the theory in the abstract, $\Delta S_k$ has the form
\be
\Delta S_{k} = \frac{1}{2}\int d^{4}x \sqrt{-g} \; \bf{\Phi} \; {\bf R}_{k} (-\Box) \; \bf{\Phi},
\ee
with the form of ${\bf R}_{k}$ suitably chosen (see below) and boldface denoting a potential matrix structure. $\Box \equiv g^{\mu \nu} {\nabla}_{\mu} {\nabla}_{\nu}$ denotes the Laplacian.

The coupling of all fields in the path integral $Z[J]$ to appropriate sources implies that all fields are treated equally and are dynamical at the quantum level. This fact consists a fundamental difference between our approach and the one suggested in Ref. \cite{Padilla:2014yea}. There, an argument was suggested for maintaining the quantum-mechanical equivalence between the unimodular theory (\ref{EH+lambda+phi2}) and GR, provided the Lagrange multiplier and $\stuck$ fields are not coupled to the sources. 
From the path integral (\ref{PI}) one can define the generating functional of connected correlators $W[J] = \ln Z[J]$, and then, through an appropriate Legendre transform of $W[J]$ the Eucledian, scale dependent (coarse--grained) effective action $\Gamma_k$. 
It can be then shown that the effective action $\Gamma_k$ satisfies an ERGE 
\cite{Wett_RG-Eq, Morris:1994ie}, 
\be
\partial_{t} \Gamma_k = \frac{1}{2} \text{Tr } \left[\left({\bf \Gamma}_{k}^{(2)} + {\bf R}_{k} \right)^{-1}  \partial_t {\bf R}_{k} \right], \label{WetterichEq}
\ee
with $k$ an infrared (IR) cut-off, $\partial_{t} \equiv k \partial/\partial k$, ``Tr" denoting a trace over momenta, tensor indices and fields, and bold-faced symbols denoting matrix quantities. What is more, ${\bf \Gamma}_{k}^{(2)}$ denotes the full, inverse propagator defined as
\be
{\bf \Gamma}_{k}^{(2)}[{\bf \Phi}] = \frac{\delta^{2} \Gamma_k}{\delta {\bf \Phi}(x) \delta {\bf \Phi}(y) }.
\ee
The regulator ${\bf R}_{k}$ ensures IR regularisation and finiteness of the trace: momentum modes below $k$ are suppressed, while those above $k$ are not, and are therefore integrated out. The form of the cut--off function is in principle arbitrary apart from some very generic requirements it has to satisfy \cite{Gies:2006wv, Reuter:2012id, Pawlowski:2005xe, Litim:2008tt} to ensure that the process of integrating out modes is performed consistently and in the Wilsonian sense, i.e shell by shell in momentum space. 

In principle, one would like to solve the ERGE (\ref{WetterichEq}) in an exact manner, but at the practical level one has to resort to the use of a (truncated) ansatz for the effective action. Given a particular ansatz for the effective action, evaluation of the ERGE (\ref{WetterichEq}) yields the flow equation, which describes the RG dynamics of $\Gamma_{k}$ as a function of the cut--off scale $k$. Our starting ansatz for the Eucledian effective action in this work will be based on the diffeomorphism invariant unimodular action presented in (\ref{EH+lambda+phi2}),
\be
\Gamma_{k}[g, \phi, \lambda ]  = - \int d^{4}x \sqrt{g} \left[Z_{G}(R - 2\Lambda)  +  \lambda f(\psi) + q(\psi) \right], \label{eff-action-UM}
\ee
where $\psi \equiv |J^\alpha{}_\beta|/\sqrt{g}$ and the $\stuck$ Jacobian defined in (\ref{Jab-def}). Noting that the (inverse) Newton's coupling acts as an overall wave-function renormalisation for the graviton, we define $Z_{G} \equiv Z_{G}(k) \equiv 1/16 \pi G(k)$. Newton's and cosmological constant are assumed to be scale-dependent renormalised couplings, i.e  $G \equiv G(k)$ and $\Lambda \equiv \Lambda(k)$ respectively. What is more, the functions $f(\psi)$ and $q(\psi)$ acquire a scale dependence through their couplings $\rho_i \equiv  \rho_{i}(k), \sigma_i \equiv \sigma_i(k)$. When solving the flow equation, we will assume for the functions $f(\psi)$ and $q(\psi)$  an expansion in powers of $\psi$ as
\be
f(\psi) = \sum_{i=0}^{N_f} \frac{1}{i!} \rho_{i}\psi^{i}, \qquad
q(\psi) = \sum_{i=1}^{N_q} \frac{1}{i!} \sigma_{i}\psi^{i}.\label{fq-ansatz}
\ee


We are now in a position to start the evaluation of the ERGE for the unimodular effective action (\ref{eff-action-UM}). We begin by calculating the inverse propagator, ${\bf \Gamma}^{(2)}_k$, from the expansion of the effective action up to second order in field fluctuations, together with the appropriate gauge fixing and ghost terms\footnote{The restriction to a particular gauge naturally introduces some dependency of the quantitative results on the gauge. For discussions of gauge invariant approaches for gauge theories within the ERG see for example \cite{Morris:1998kz,Morris:1999px,Morris:2000fs,Arnone:2000qd,Arnone:2006ie}.} (see below.) In this regard, we first split the fields $g_{\mu \nu}(x), \, \lambda(x), \,  \phi^{\alpha}(x)$ into a background and a perturbation piece as
\begin{align}
& g_{\mu \nu}(x) = \bar{g}_{\mu \nu}(x) + G_0^{1/2}h_{\mu \nu}(x), \nonumber \\
& \lambda(x) = \bar{\lambda}(x) + G_0^{-1/2} \delta \lambda(x), \nonumber \\
& \phi^{\alpha}(x) = \bar{\phi}^{\alpha}(x) + G_0^{1/2} \delta \phi^{\alpha}(x),
\end{align}
with $G_0=G_N/8 \pi$ related to Newton's constant, $G_N$, as it is measured today. The metric and $\stuck$ fields are dimensionless and the Lagrange multiplier $\lambda$ is of mass dimension two, while all fields' fluctuations are normalised to mass dimension equal to one. To diagonalise the inverse propagator we split the metric perturbation into the trace and trace-free parts $h_{\alpha \beta} = \hat{h}_{\alpha \beta} + \frac{1}{4}\bar{g}_{\alpha \beta}h$  ($\bar{g}_{\alpha \beta} \hat{h}^{\alpha \beta} = 0$), and the $\stuck$ fluctuation into the transverse and longitudinal parts $ \delta \phi^{\alpha} = \delta \hat{\phi}^{\alpha}  +  G_{0}^{1/2} \bar \nabla^{\alpha} \delta \phi$  ($ \bar \nabla_{\alpha} \delta \hat{\phi}^{\alpha} = 0$).

We choose to work with the de Donder gauge, which accounts to the addition of the following gauge fixing term in the effective action
\be
 S_{\text{GF}} = Z_{G}\int d^{4}x \sqrt{\bar{g}} \bar{g}^{\mu \nu} h_{\alpha \beta}  \mathcal{F}_{\mu}^{\alpha \beta} \mathcal{F}_{\nu}^{\gamma \delta} h_{\gamma \delta}, \label{GF-de-Donder} 
\ee
\be
 \mathcal{F}_{\mu}^{\alpha \beta} \equiv \delta^{\beta}_{\mu} \bar{g}^{\alpha \gamma} \bar{\nabla}_{\gamma} - \frac{1}{2}  \bar{g}^{\alpha \beta}\bar{\nabla}_{\mu}.
\ee
		
What is more, there are two ghost contributions resulting from the corresponding determinants in the path integral volume, $S^{\text{(ghost)}} = S_{\text{EH}}^{\text{(ghost)}} + S_{\text{St\"uck.}}^{\text{(ghost)}}$. The first contribution comes from the gauge fixing term (\ref{GF-de-Donder}) which introduces a determinant in the path integral measure $\sim | \mathcal{F}_{\mu}^{\alpha \beta}|$, which in turn can be represented in terms of Grassmann fields according to the Fadeev--Poppov procedure as \cite{Reuter:1996cp}
\be
S_{\text{EH}}^{\text{(ghost)}} = - \int d^{4}x\sqrt{\bar{g}} \, \, \bar{C}_{\mu} \left(  - \delta^{\mu}_{\nu}\Box  - \frac{1}{4}\delta^{\mu}_{\nu}R  \right) C^{\nu}, 
\ee
with $C_\nu$ and $\bar{C}_{\nu}$  the ghost and anti-ghost fields respectively. From now on, $\Box \equiv \bar{g}^{\mu \nu} \bar{\nabla}_{\mu} \bar{\nabla}_{\nu}$ will be denoting the Laplacian with respect to the background metric.

In a similar fashion, the St\"uckelberg field decomposition into a transverse and longitudinal part introduces a non--trivial determinant which can be also exponentiated through the Fadeev--Poppov recipe, contributing the following term in the effective action (see Appendix \ref{ghost} for an explicit derivation)
\be
S_{\text{St\"uck.}}^{\text{(ghost)}} \equiv - \frac{1}{2}\int d^{4}x\sqrt{\bar{g}} \, \,  \bar{\eta} (- \Box) \eta.
\ee
Notice that for both ghost actions we have assumed a non-running wave--function renormalisation, and we have essentially set $Z_{\text{St\"uck.}}^{\text{(ghost)}} = Z_{\text{EH}}^{\text{(ghost)}} = 1$. The expanded effective action supplemented with the appropriate gauge fixing and ghost terms reads schematically
\begin{align}
& { \Gamma}_{k}^{(2)} + S_{\text{GF}}+ S^{\text{(ghost)}} = 
 \int d^{4}x \sqrt{\bar g}\Bigg\{ \hat{h}_{\mu \nu} \left[ \Gamma^{(2)}_{\hat{h}\hat{h}}\right]^{\mu \nu}_{\rho \sigma}   \hat{h}^{\rho \sigma} \no \\
%
&+  \frac{1}{4}  h  \left[\Gamma^{(2)}_{hh}\right]   h + h \left[ \Gamma^{(2)}_{h \lambda}\right] \delta \lambda +  \delta \lambda \left[ \Gamma^{(2)}_{\lambda \phi} \right] \delta \phi \no \\
%
 & +  \delta \phi \left[\Gamma^{(2)}_{\phi \phi }\right] \delta \phi  +  h \left[\Gamma^{(2)}_{h\phi}\right]  \delta \phi  + \delta \hat{\phi}^{\mu} \left[\Gamma^{(2)}_{\hat{\phi} \hat{\phi}}\right]_{\mu \nu} \delta \hat{\phi}^{\nu}  \Bigg\},  \label{eff-action-second-full} 
\end{align}
with the cut--off dependent second functional derivatives of the corresponding fields, $\Gamma^{(2)}_{AB}$, presented explicitly in equation (\ref{Hessian-entries}) of the appendix.  

We will be restricting ourselves to a background described by a Euclidean sphere $S_4$, which will also allow us to use the method of the heat kernel expansion for the evaluation of trace integrals, see Appendix \ref{trace}. With this choice, the background Riemann tensor can be expressed as \footnote{For a recent explicit analysis of various subtleties of the background field method within the ERG of scalar field theories see Ref. \cite{Bridle:2013sra}, while for recent developments for the case of metric gravity Ref. \cite{Becker:2014qya} respectively. In the same context, the dependency of the background topology within the $f(R)$ approximation has been recently studied in \cite{Demmel:2014sga}. }$
\bar{R}_{\alpha \beta \gamma \delta} = \frac{1}{12}\bar{R} \left( \bar{g}_{\alpha \gamma}\bar{g}_{\beta \delta} - \bar{g}_{\beta \gamma}\bar{g}_{\alpha \delta} \right)
$.
For the background value of the St\"uckelberg scalars we take $\bar{\phi}^{\alpha} = \epsilon x^{\alpha}$ 
with $\epsilon$ an arbitrary real number, and for the Lagrange multiplier we take $\lambda=\bar \lambda = \text{constant}$. This choice will in addition lead to equations which are easier to deal with. 
In what follows, we will drop the bar over background fields for brevity.  

We now turn to the scale dependent cut--off functions ${ \bf R }_{k}$, which will modify the inverse propagators in the ERG equation as $ {\bf \Gamma}^{(2)}_{k}  + { \bf R }_{k}$. Notice that ${ \bf R }_{k}$ should carry the same overall index structure with the corresponding Hessian entry, so that it combines appropriately. We will use a type Ia cut-off, which modifies the effective action at quadratic order (\ref{eff-action-second-full}) by trading each Laplacian as $-\Box \to   -\Box + R_{k}(-\Box)$ \cite{Reuter:1996cp}. The cut--off $R_{k}(-\Box)$ is constructed such that it cuts off eigenvalues of the Laplacian with value less than $k$, implementing this way an infrared regularisation. For our profile function $R_{k}(-\Box)$ we will make a choice that will allow us to evaluate the trace integrals in closed form \footnote{For an investigation of the cut--off dependency of the gravitational beta functions within the ERG see for example Ref. \cite{Narain:2009qa}.}; we choose Litim's ''optimised" cut--off $R_{k}(-\Box) = (k^2 + \Box)\Theta(k^2 + \Box)$ \cite{Litim_Opt-CO1}, with $\Theta$ the Heaviside step function. 

In the evaluation of the trace on r.h.s of the ERGE we will set to zero those regulators $R_{AB}$ which correspond to momentum independent interactions, i.e those which do not involve any Laplacian $-\Box$. Looking at the explicit expressions (\ref{Hessian-entries}) we set for the non--propagating interactions, $R_{\hat{\phi}\hat{\phi}}= 0, \; \; R_{h \lambda} = 0$.


The l.h.s of the ERGE, (\ref{WetterichEq}), can be evaluated in a straightforward way as
 \begin{align}
 \partial_{t}\Gamma_{k} =  \tilde{\text{Vol}}  \Big[ 2(\eta_{\Lambda} + & \eta_{G})\tilde{Z}_{G}\tilde{\Lambda} - \eta_{G} \tilde{Z}_{G} \tilde{{R}} 
 - \eta_{f} f(\psi) \tilde{{\lambda}} - \eta_{q} q(\psi) \Big], \label{ERGE-lhs}
\end{align}
where $\tilde Z_G=1/16\pi \tilde G$ and $\tilde{\text{Vol}}=\frac83 \frac{\pi^2}{\tilde R^{4}}$, the latter  corresponding to the volume of the 4-dimensional sphere in units of the cut-off.
The anomalous dimensions associated with $G$, $\Lambda$ and $f, q$ respectively are defined as
\be
\eta_{G} \equiv \frac{ \partial_{t} Z_{G}}{Z_G}, \; \; \; \;  \eta_{\Lambda} \equiv \frac{\partial_{t} \Lambda}{\Lambda}, \; \; \; \;  \eta_{X} \equiv \frac{ \partial_{t} X}{X},
\ee
for $X= \{ f, f', f'', q, q', q''\}$, 
while the tildes denote dimensionless quantities with respect to the cut--off $k$, i.e. 
\be
 \tilde{G} \equiv  \tilde{G}(k) \equiv k^2 G(k),  \; \; \tilde{\Lambda} \equiv \frac{\Lambda(k)}{ k^{2}},  \; \;  \tilde{R}\equiv \frac{{\bar R}}{k^{2}}, \;\; \tilde \lambda =\frac{\bar \lambda}{k}.
\ee
Finally, on our choice of  background, the field $\psi$ acquires the value $\bar{\psi} = \epsilon^4/\sqrt{\bar{g}}$. 

We are now in a position to put all pieces together into the r.h.s of the ERGE, and evaluate the trace over fields and momenta. That would then allow us to extract the running of the effective action by combining the result with relation (\ref{ERGE-lhs}). We perform the momentum integration in the trace integral asymptotically, using a small scale heat kernel expansion, where $\bar{R}/k^2 \ll1$ (see appendix \ref{trace} for more details.) After the trace evaluation, the contributions on the r.h.s of the ERGE (\ref{WetterichEq}) coming from the tensor, scalar and ghost fluctuations read as
\begin{align}
\left( \tilde{{\text Vol}} \right)^{-1} \partial_{t} \Gamma_{k} =   \left. \partial_{t} \Gamma_{k} \right |_{ \text{T}}     +    \left. \partial_{t} \Gamma_{k} \right |_{\text{Scalar}}   +   \left. \partial_{t} \Gamma_{k} \right |_{\text{Ghosts}}, \label{ERGE-rhs}
\end{align}
with 
\begin{align}
\left. \partial_{t} \Gamma_{k} \right |_{\text{Ghosts}} = -\frac{5 \tilde{R}+9}{32 \pi ^2}, \label{Ghost-contribution}
\end{align}
\begin{align}
 \left. \partial_{t} \Gamma_{k} \right |_{\text{T}}   = \frac{9 \left((\tilde{R}+2) \eta _G+4 (\tilde{R}+3)\right) \tilde{Z}_G}{128 \pi ^2 \left(3 \tilde{U}+3 \tilde{\lambda}  \tilde{V}+\tilde{Z}_G (2 \tilde{R}-6 \tilde{\Lambda}+3)\right)}, \label{Tensor-contribution}
\end{align}

for the ghost and tensor parts respectively. The scalar sector, resulting from the fluctuations of the fields $h$ and $\phi$, is more involved. Defining $' \equiv d/d\psi$, and {\it omitting the tildes above all dimension full quantities (e.g $R, f, q$) only for this case} in order to simplify the notation, the scalar contribution can be expressed as
\begin{equation}
\left. \partial_{t} \Gamma_{k} \right |_{\text{Scalar}} = \frac{S_n}{S_d},
\end{equation}
with
\begin{widetext}
\begin{align}
& S_{n} = V^2  \Big[    
 R^2 \left( 3 \left(\lambda  \left(\eta_{f'}+4\right) f'+\left(\eta _{q'}+4\right) q'\right) \right) 
 + 2R \left(3 \lambda  \left(\eta_{f'}+6\right) f'-8 \lambda  \psi  \left(\eta_{f''}-6\right) f''+3 \eta_{q'} q'+18 q'-8 \psi  \eta_{q''} q''+48 \psi  q''\right) \no \\  
&- 36 \psi  \left(\lambda  \left(\eta _{f''}-8\right) f''+\left(\eta _{q''}-8\right) q''\right) 
\Big] 
 + 12 \psi f' V  \Big[     
 R  \left(2 \psi  \left(\lambda  \left(\eta_{f'}+\eta _{f''}+8\right) f''+\left(\eta_{f'}+\eta _{q''}+8\right) q''\right)+\lambda  \left(\eta_{f'}+4\right) f'\right)      \no \\  
&  + 2\left(2 \psi  \left(\lambda  \left(\eta_{f'}+\eta _{f''}+12\right) f''+\left(\eta_{f'}+\eta _{q''}+12\right) q''\right)+\lambda  \left(\eta_{f'}+6\right) f'\right)
  \Big]  \no \\ 
& +  6 \psi  f'^2 \Big[  
       R \left(4 \lambda  \psi ^2 \left(\eta_{f'}+4\right) f''+4 \psi ^2 \left(\eta_{f'}+4\right)
      q''+2 \eta_{f'} U+8 U  +  2 \eta_{f'} Z_G(1 - 2 \Lambda) +  \eta _G Z_G + 4Z_G (3 - 4\Lambda)  \right)    \no \\  
& + 2 \left(4 \lambda  \psi ^2 \left(\eta_{f'}+6\right) f''+4 \psi ^2 \left(\eta_{f'}+6\right)
   q''+2 \eta_{f'} U+12 U  + 2 \eta_{f'} Z_G(1 - 2 \Lambda) + \eta _G Z_G   +  Z_G (3 - 4\Lambda) \right) \Big]  \label{Scalar-num}\\
%
%
& S_{d} = 1152 \pi ^2 \Big[ (\lambda  R f'+4 \lambda  \psi  f''+R q'+4 \psi  q'') 
+ 2 \psi  f' (\lambda  f'+4 \psi  (\lambda  f''+q'')) 
 + 2 \psi  f'^2 \left(2 \lambda  \psi ^2 f''+2 \psi ^2 q''+U +Z_G(1 - 2\Lambda) \right) \Big]. \label{Scalar-denom}
\end{align}
\end{widetext}

Equation (\ref{ERGE-rhs}), combined with (\ref{ERGE-lhs}),  is the first main result of this work. It corresponds to the (non--perturbative) RG flow equation of the effective action (\ref{eff-action-UM}), under the particular assumptions of background and gauge stated earlier. In particular, it describes how the couplings $\tilde G, \tilde \Lambda$ as well as $\tilde f, \tilde q$ flow under a change of the cut--off scale $k$. Notice that the RG derivatives, $\partial_t$, appear on both sides of the flow equation in a rather complicated way through the  quantities $\eta_i$. Solving the flow equation without a particular assumption about the form of the functions $f$ and $q$ is, if possible at all, a very involved task. Here,  we will attempt to solve it using the method of truncation,assuming a polynomial expansion for $f$ and $q$, as presented in (\ref{fq-ansatz}). 

We now set Eq. (\ref{ERGE-lhs}) equal to Eq. (\ref{ERGE-rhs}), and expand both sides order by order in the background fields, $\tilde R, \frac{\epsilon^4}{\sqrt{\bar{g}}}$ and $\tilde \lambda$. This gives a number of equations from which we can extract the beta functions for the running of the dimensionless couplings as functions of the dimensionless couplings themselves,
\begin{align} 
\partial_t \tilde c =\beta_{\tilde c}(\tilde G, \tilde \Lambda, \tilde \rho_i, \tilde \sigma_i) \equiv \tilde c \left( -d + \eta_{\tilde c}(\tilde G, \tilde \Lambda, \tilde \rho_i, \tilde \sigma_i) \right)  ,
\end{align}
with $\tilde c = \{\tilde \Lambda, \tilde G, \tilde \rho_j, \tilde \sigma_j \}$, $d$ the canonical mass dimension and $\eta_{\tilde c}$ the associated anomalous dimension of the corresponding coupling. The canonical dimension is the trivial contribution expected from usual power counting arguments, while the anomalous one is non--trivial and takes into account the effect of (off--shell) quantum corrections to the running of the coupling. 

Since the final step in our analysis will be to compare the quantum structure of the unimodular action (\ref{eff-action-UM}) with that of standard GR, it is important to first define which will be the properties we will be comparing. Given the two theories and the associated RG flows, we shall be comparing the following properties:
\begin{itemize}[leftmargin=*]
\item The UV-fixed point structure under the RG, i.e the asymptotic values of the (dimensionless) couplings of the theory as the cut--off scale $k$ is taken to infinity. 
\item The structure of the linearised RG flow around the corresponding fixed points, or in other words the structure of the critical manifold. In particular, we will be interested in the nature of the associated eigendirections (relevant/irrelevant), as well as the dimensionality of the critical manifold. 
\end{itemize}

In this sense, we will be comparing the UV completion of the two theories. 
Our strategy will be to calculate the above quantities first for the case of standard GR, and then proceed by turning on the interactions in the $\stuck$ sector order by order.

Before we proceed with our analysis, let us review some useful concepts from the RG and asymptotic safety. The scenario of asymptotic safety in the context of GR was first suggested by Weinberg 
\cite{Weinberg:1980gg} 
\footnote{For reviews on the RG and asymptotic safety see Refs. \cite{Morris:1998da, Gies:2006wv, Niedermaier:2006wt, Percacci:2007sz, Litim:2008tt, Reuter:2012id, Reuter:2012xf}, while for applications in a cosmological setting \cite{Wein, Reuter:2005kb,Hindmarsh:2012rc,Contillo:2011ag,Cai:2011kd,Hindmarsh:2011hx,Bonanno:2001xi,Reuter:2005kb, Reuter:2012xf,Copeland:2013vva}.}, 
as a non--perturbative UV completion of gravity. For metric theories of gravity, there have been significant indications for the existence of an appropriate UV fixed point associated with a three dimensional critical surface under the RG for GR, as well as for actions with higher powers of curvature
\cite{Reuter:1996cp, Falkenberg:1996bq, Souma:1999at, Lauscher:2001ya,Lauscher:2002sq, Litim:2003vp, Reuter:2001ag, Dou:1997fg, 
Codello:2007bd,Codello:2008vh,Machado:2007ea,Benedetti:2009gn,Benedetti:2009rx,BenedettiCaravelli, Benedetti:2011ct, 
Falls:2013bv,Manrique:2011jc,Eichhorn:2013ug,Demmel:2013myx,Dietz:2012ic,Dietz:2013sba,Falls:2014zba,Falls:2014tra}. In the context of asymptotic safety, renormalisation is achieved if the essential couplings of the theory approach asymptotically a fixed point under the RG, as the cut--offf scale $k$ is taken to infinity. The fixed point, which we denote as  $\tilde g=\tilde g_*$, coresponds to the vanishing of the associated beta function $\beta_{\tilde g_i}(\tilde g_*)=0$. A crucial point is that in the context  of asymptotic safety, the smallness of the coupling(s) in the fixed point regime is not required, an immediate consequence of the non--perturbative character of the scenario. For the theory to be predictive, there have to be only a finite number of relevant (attractive) couplings in the UV, or in other owrds, that the associated critical manifold is finite dimensional. The latter requirement implies that only a finite number of (relevant) couplings will have to be fixed in a given experiment. 
The attractivity properties of the couplings can be studied through the linearised flow around a given fixed point, which is mathematically expressed by the linearisation matrix
\be
\partial_{t}\tilde{c}_{i} = \sum_{j} \left. \frac{\partial \beta_{i}(\tilde{c}_n)}{\partial \tilde{c}_j} \right|_{\tilde{c}_j = \tilde{c}_{j*}} \times (\tilde{c}_j -  \tilde{c}_{j*}), \label{linear-matrix}
\ee
with $\tilde{c}_j$ denoting the couplings of the theory under study measured in units of the cut--off $k$. The number of relevant directions corresponds to the number of negative eigenvalues of the matrix $\frac{\partial \beta_{i}(\tilde{c}_n)}{\partial \tilde{c}_j}$.
\subsection{Minimal unimodular case}

Let us now start with the general unimodular action, and first consider what happens in the two most simple cases: the Einstein-Hilbert truncation ($f=q=0$), and the minimal unimodular truncation ($f=\rho_0, ~q=0$).  The former corresponds to the case of standard GR, while the latter to a unimodular action where the $\stuck$ interactions are kept minimal.

For the Einstein-Hilbert truncation, i.e with the $\stuck$ sector completely switched off from the onset, we find a Gaussian fixed point (GFP) at $(\tilde{\Lambda}, \tilde{G}) = (0, 0)$, and a single, attractive UV fixed point at $(\tilde{\Lambda}, \tilde{G}) =(0.193, 0.707)$ with eigenvalues $(\gamma_{\Lambda}, \gamma_{G}) \simeq \left( -1.99 \pm 3.829 i  \right)$, which confirms previous findings in the literature \cite{Cod-Perc-Rahme1}. The negative real part of the eigenvalues implies that they correspond to relevant eigendirections along which the couplings are attracted towards the fixed point. The GFP is a trivial fixed point and is the one around which perturbation theory is usually applied, i.e the small coupling regime. On the other hand, the non-trivial UV fixed point corresponds to the limiting value of the couplings as $k \to \infty$. 

We now turn on the $\stuck$ sector in a minimal way, by setting in (\ref{fq-ansatz})
\be
q(\psi) = 0, \; \; f(\psi) = \rho_0 + \rho_{1} \psi. 
\ee
For this case, we find that the UV fixed point for the cosmological and Newton's coupling persists and it equals to
\be
(\tilde{\Lambda}, \tilde{G}) =(0.206, 0.661), \label{GLFP-minimal}
\ee 
while the two $\stuck$ couplings of the unimodular sector become trivial, i.e 
\be
\tilde{\rho}_0 = \tilde{\rho}_1 = 0.  \label{Rho-sigmaFP-minimal}
\ee  
The stability analysis around the fixed point (\ref{GLFP-minimal})--(\ref{Rho-sigmaFP-minimal}) reveals that all couplings are relevant with eigenvalues $\gamma_{\tilde{\Lambda}, \tilde{G}} = -1.611\pm 3.244i$ and $\gamma_{\tilde{\rho}_0} = -6.066$ and $\gamma_{\tilde{\rho}_1} = -2$ respectively. Notice that the fixed point structure of the Einstein-Hilbert--sector (fixed points and associated eigenvalues) around the fixed point are qualitatively the same as for those of standard GR. The actual numerical values of the associated fixed point and eigenvalues are also very close for both theories. It is very interesting to notice also that for the dimensionless product $G \times  \Lambda$ evaluated on the UV fixed point we have that $\left(  \tilde{G}_{*}\tilde{\Lambda}_{*} \right)_{\text{Unimod.}} \simeq  \left( \tilde{G}_{*}\tilde{\Lambda}_{*} \right)_{\text{GR}} \simeq 0.136$. 
What is more, we also notice the non--trivial value of $\gamma_{\tilde{\rho}_0}$, which as we shall see later will be a more general feature. 
The results for higher truncations in the $\stuck$ sector are summarised in Table \ref{maintable}. 

We conclude that in this context, the minimal unimodular case shares essentially the same UV features with its GR counterpart. As well shall see later, the behaviour found in the minimal case persists to higher orders in the $\stuck$ sector; the cosmological and Newton's coupling run to an attractive non--trivial UV fixed point similar to that of standard GR, while the $\stuck$ couplings remain trivial. 

\subsection{Higher order $\stuck$ sector}

We now discuss the extension of the previous results to the case where we include higher order terms in the $\stuck$ sector through the expansion (\ref{fq-ansatz}). In particular, we consider three different cases; the case where $q = 0$ and $f \neq 0$ with $N_f = 4$, then a second case where $f = 0$ and $q \neq 0$ with $N_q = 4$, and finally the third case where $f, q \neq 0$ with $N_f = 1$ and $N_q = 2$ respectively. Notice that, for the case with $f=0$, the expansion in $q$ has to start at least second order in $\psi$ for the classical equations of motion to be consistent, as explained in Ref. \cite{Padilla:2014yea}. For each case, we proceed by increasing the interactions in the $\stuck$ sector step by step. The results, which are summarised in Table \ref{maintable}, reveal that the attractive, UV fixed point for the Einstein--Hilbert sector persists and is stable, while the  number of relevant couplings in the St\"uckelberg sector increases with the order of the truncation, the latter issue endagering the predictive power of the theory. We will explain later how this issue could be resolved. 

For the first case, i.e $q = 0$ and $f \neq 0$ with $N_f = 4$, the truncation with $N_f =2$ corresponds to the minimal unimodular case discussed in the previous subsection. For $N_f > 2$, more than one potentially acceptable fixed points for $\tilde \Lambda, \tilde G$ appear, depending on the way the $\stuck$ couplings $\tilde{\rho}_i$ approach zero in the UV, a behaviour which is a direct result of the structure of the corresponding beta functions. To understand this better, let us consider as an example the case with $N_f = 3$ ($q = 0$). The beta functions of $\tilde \Lambda$, for example, acquire schematically the form
\be
\beta_{\tilde \Lambda} = -2 \tilde \Lambda   +  \frac{   \sum_{n = 0}^{4} a_{n}(\tilde \Lambda, \tilde G) \tilde \rho_{0}^n \tilde \rho_{1}^{8-2n} \tilde \rho_{2}^n     }{\sum_{ m= 0}^{4} b_m(\tilde \Lambda, \tilde G) \tilde \rho_{0}^m \tilde \rho_{1}^{8-2m} \tilde \rho_{2}^m },
\ee
with $a_{n}, b_{m}$ non--trivial functions of $\tilde \Lambda$ and $\tilde G$. $\beta_{\tilde G}$ acquires a similar form with the above, with a different set of $a_{n}, b_{m}$. Notice that the second term of $\beta_{\tilde \Lambda}$ effectively corresponds to the anomalous dimension associated with the coupling $\tilde \Lambda$.
 Now, considering the limit of either $\tilde \rho_{0} \to 0$ or $\tilde \rho_{2} \to 0$ in the above equation, the fraction on the r.h.s gives $a_{1}/b_{1}$, while the limit $\tilde \rho_{1} \to 0$ gives $a_5/b_5$ respectively. Obviously, these two cases, lead to two different fixed point equations for $\tilde \Lambda, \tilde G$. We will not be interested in discussing explicitly this behaviour here, and we leave it for a possible future work. 
Our criterion for selecting out only one of them has been the existence of the correct IR limit, i.e that the UV fixed point is viably connected with the IR regime along the RG flow, an issue we explored numerically for a sample of initial conditions. \footnote{Of course, we would not like to claim by any means that this is a conclusive statement, as an exhaustive numerical analysis would be needed. However, that would be far beyond the scope of this work.}

The other two truncation  ans\"atze we considered, i.e with $f = 0, q \neq 0$ and $f, q, \neq 0$ exhibit similar behaviour as the first case described above, with all the corresponding fixed point and eigenvalues respectively appearing to be stable and attractive (see also Table \ref{maintable}.) In particular, the fixed-point and eigenvalues of the Einstein--Hilbert sector appear to be stable with respect to the previous cases, while the St\"uckelberg-sector couplings remain trivial and attractive.

It is interesting to notice that the dimensionless product $\tilde{G}_{*}\tilde{\Lambda}_{*}$ for all unimodular cases agrees with that of GR up two decimal places. In particular, for all the unimodular truncations presented in Table \ref{maintable}, as well as the case of GR we have that,
\begin{align}
& \left.  \tilde{G}_{*}\tilde{\Lambda}_{*} \right|_{\text{Unimod.}} \simeq  \left. \tilde{G}_{*}\tilde{\Lambda}_{*} \right|_{\text{GR}} \simeq 0.13.
\end{align} 

Another important point which is common in all truncations we studied is the fact that in the St\"uckelberg sector, the number of relevant (attractive) couplings increases with the order of truncation (see also Table \ref{maintable}), a behaviour which is in contrast to the requirements for an asymptotically safe theory, since in that case the dimensionality of the critical manifold increases with the truncation and the theory becomes non--predictive. This potentially problematic behaviour has its root in the fact that $\psi$ carries mass dimension zero already in the classical unimodular action. Remember that, in the context of a standard scalar field theory with a canonical kinetic term and a potential, the scalar couplings, at least at power counting level, are relevant up to order four in powers of the scalar self-interactions. However, as it turns out, for the dimensionless scalar $\psi$ considered in this work, the non--trivial quantum interactions in the UV cannot win against the canonical mass dimensions of the couplings $\rho$ and $\sigma$ respectively, leading to UV-attractive scalar couplings only. Let us see what would happen if in the spirit of usual scalar field theories we canonically normalise the field $\psi$, by making the fied redefinition $\psi \to \psi/Z_{\psi}^{1/2}$, with $Z_{\psi}$ a wave-function renormalisation of mass dimension two, which we assume to be independent of the cut--off scale  \footnote{The author is thankful to Astrid Eichhorn for bringing this possibility to his attention, as well as for useful discussions around this point.}.  The effect of the field redefinition would be to shift the canonical mass dimension of the St\"uckelberg couplings $\rho_n$ and $\sigma_n$ by $n$. For example, including a kinetic term in the effective action, i.e $Z_{\psi} (\partial \psi)^2$, and requiring that it is canonically normalised, would lead to a redefinition of $\psi$ as explained before, and in fact, it is natural to expect that in principle quantum effects would generate such a kinetic term. 
To see therefore how the situation would change according to the above discussion, one would have to shift all of the St\"uckelberg eigenvalues in Table \ref{maintable} by adding $n$, i.e 
\begin{align}
\gamma_{\tilde{\rho}_n} \to \gamma_{\tilde{\rho}_n}+n, 
\end{align}
with a similar shift for $\gamma_{\tilde{\sigma}_n}$.
It is then straightforward to see that after doing this, and for all truncations considered, there is a point after which the St\"uckelberg couplings become irrelevant, or in other words the associated eigenvalues become positive. For example, for the truncation with $q = 0$ and $f \neq 0$ with $N_f = 4$,  the couplings $\rho_n$ with $n\geq 2$ become irrelevant (eigenvalues $>0$), while those with $n<2$ remain relevant (eigenvalues $<0$). Notice that, even if we had included a kinetic term for $\psi$ in the effective action from the onset, that would not be expected to affect any of our conclusions, apart from the nature of the St\"uckelberg-sector eigenvalues as discussed before. 


{The similarity found between the unimodular theory and GR at sufficiently high energies should not be regarded as any sort of general statement about the equivalence of the two theories at the quantum level. In fact, although the $\stuck$ couplings turn off in the deep UV regime ($k \to \infty$), as the cut--off scale $k$ is lowered towards the IR the RG flow will in principle drive them  away from zero, leading to non--trivial interactions in the $\stuck$ sector of the theory. As a result, the non--trivial interactions in the $\stuck$ sector will lead to different structure for the respective effective actions of the two theories, as well as to different phenomenologies at the quantum level. This should come as no surprise, since by construction the Lagrange multiplier and $\stuck$ fields were allowed to couple to appropriate sources in the generating functional (\ref{PI}).
}

Before we conclude this section, let us comment on the non--trivial eigenvalues associated with the couplings $\tilde{\rho}$ and $\tilde{\sigma}$, different from the corresponding mass dimension as one would expect due to the triviality of these couplings. In fact, a similar feature has been also observed before in the literature, and in particular within the context of scalar--tensor theories, see for example Refs \cite{Narain:2009fy,Narain:2009gb}. The reason behind its occurrence is the presence of terms linear with respect to these couplings in the corresponding beta functions. 

We conclude that for the unimodular theory, a non-trivial and attractive UV fixed point for the Einstein--Hilbert sector, spanned by the couplings $G, \Lambda$, always exists, while the $\stuck$ couplings become trivial. In this sense, the unimodular effective action shares similar features with the corresponding one of GR in the UV: since the $\stuck$ sector of the unimodular theory turns off at sufficiently high energies, while the Einstein--Hilbert sector exhibits essentially the same fixed point structure with that of standard GR, the associated effective actions acquire similar forms. In particular, the fixed point values for the couplings $G, \Lambda$ in both theories were found to be similar, and the same was true for the associated (complex conjugate) eigenvalues. Our analysis further revealed that the attractivity properties of the $\stuck$ sector are endangered by the dimensionless nature of the field $\psi$ in the original unimodular action, leading to a critical manifold which' dimensionality increases with the order of the truncation in the $\stuck$ sector. Canonically normalising the field suggests that the theory becomes predictive, with a critical manifold which is finite.

\begin{table*} 
\begin{tabular}{ |l |l |l |l |l |l |l |l |l |l |l | }
\hline
   $\tilde{\Lambda}_{*}$ & $\tilde{G}_{*}$ & $\tilde{\rho}_{0*}$ & $\tilde{\rho}_{1*}$ & $\tilde{\rho}_{2*}$ & $\tilde{\rho}_{3*}$ & $\tilde{\rho}_{4*}$  & $\tilde{\sigma}_{1*}$  & $\tilde{\sigma}_{2*}$ & $\tilde{\sigma}_{3*}$ & $\tilde{\sigma}_{4*}$   \\
   \hline
      $0.193$ & $0.707$        &  $\;\;\;\; -$    & $\;\;\;\; -$     & $\;\;\;\; -$   &  $\;\;\;\; -$  &  $\;\;\;\; -$ &  $\;\;\;\; -$  & $\;\;\;\; -$   &  $\;\;\;\; -$  &  $\;\;\;\; -$  \\
        \hline 
         \hline
      $0.206$ & $0.661$        &  $0$    & $0$     & $\;\;\;\; -$   &  $\;\;\;\; -$  &  $\;\;\;\; -$ &  $\;\;\;\; -$ &  $\;\;\;\; -$   &     $\;\;\;\; -$   &     $\;\;\;\; -$   \\
      
          \hline 
     $0.206$ & $0.661$        &  $0$    & $0$     & $0$   &  $\;\;\;\; -$  &  $\;\;\;\; -$ &  $\;\;\;\; -$ &  $\;\;\;\; -$   &     $\;\;\;\; -$   &     $\;\;\;\; -$   \\

          \hline 
       $0.206$ & $0.661$       &  $0$    & $0$     & $0$   &  $0$  &  $\;\;\;\; -$ &  $\;\;\;\; -$ &  $\;\;\;\; -$   &     $\;\;\;\; -$   &     $\;\;\;\; -$   \\

          \hline 
       $0.206$ & $0.661$       &  $0$    & $0$     & $0$   &  $0$  &  $0$ &  $\;\;\;\; -$ &  $\;\;\;\; -$   &     $\;\;\;\; -$   &     $\;\;\;\; -$   \\
    \hline
    \hline
       $0.201$ & $0.674$        &  $\;\;\;\; -$  &     $\;\;\;\; -$   &     $\;\;\;\; -$   &     $\;\;\;\; -$   & $\;\;\;\; -$     & $\;\;\;\; -$   &  ${0}$  &  $\;\;\;\; -$ &  $\;\;\;\; -$  \\
      
       \hline 
     $0.202$ & $0.670$        &  $\;\;\;\; -$  &     $\;\;\;\; -$   &     $\;\;\;\; -$   &     $\;\;\;\; -$   & $\;\;\;\; -$     &  $\;\;\;\; -$   &  ${0}$  &  ${0}$ &  $\;\;\;\; -$  \\
      
       \hline 
     $0.203$ & $0.666$        &  $\;\;\;\; -$  &     $\;\;\;\; -$   &     $\;\;\;\; -$   &     $\;\;\;\; -$   & $\;\;\;\; -$     &  $\;\;\;\; -$   &  ${0}$  &  ${0}$ &  ${0}$  \\
   \hline   
  \hline 
      $0.206$ & $0.661$        &  $0$  &     $0$   &     $\;\;\;\; -$   &     $\;\;\;\; -$   & $\;\;\;\; -$     & $0$   &  $\;\;\;\; -$  &  $\;\;\;\; -$ &  $\;\;\;\; -$  \\
      
       \hline 
       $0.206$ & $0.661$        &  $0$  &     $0$   &     $\;\;\;\; -$   &     $\;\;\;\; -$   & $\;\;\;\; -$     & $0$   &  ${0}$  &  $\;\;\;\; -$ &  $\;\;\;\; -$  \\

  \hline
   \hline
    $\gamma_{\tilde{\Lambda}}$ & $\gamma_{\tilde{G}}$  & $\gamma_{\tilde{\rho}_0}$  & $\gamma_{\tilde{\rho}_1}$ & $\gamma_{\tilde{\rho}_2}$  & $\gamma_{\tilde{\rho}_3}$ & $\gamma_{\tilde{\rho}_4}$ &  $\gamma_{\tilde{\sigma}_1}$ & $\gamma_{\tilde{\sigma}_2}$& $\gamma_{\tilde{\sigma}_3}$ & $\gamma_{\tilde{\sigma}_4}$    \\
     \hline
    $-1.475+3.043i$         & $-1.475-3.043i$                      & $\;\;\;\;{-}$ &            $\;\;\;\;{-}$&              $\;\;\;\;{-}$ &         $\;\;\;\;{-}$&             $\;\;\;\;{-}$ &             $\;\;\;\;{-}$ &              $\;\;\;\;{-}$&      $\;\;\;\;{-}$   &        \\
     \hline
      \hline
    $-1.611 + 3.244i$         &     $-1.611 - 3.244i$                & $-6.066$ &      $-2$&             $-$ &          $\;\;\;\;{-}$&             $\;\;\;\;{-}$&               $\;\;\;\;{-}$&              $\;\;\;\;{-}$&        $\;\;\;\; -$   &       \\
    
      \hline
  $-1.611 + 3.244i$         &     $-1.611 - 3.244i$              & $-6.066$ &      $-2$&             $-0.780$ &           $-$&         $\;\;\;\; -$&          $\;\;\;\; -$&             $\;\;\;\; -$&        $\;\;\;\; -$   &       \\
    
       \hline
      $-1.611 + 3.244i$         &     $-1.611 - 3.244i$              & $-6.066$ &      $-2$&            $-0.780$ &          $-1.458$  &      $\;\;\;\; -$&         $\;\;\;\; -$&            $\;\;\;\; -$&         $\;\;\;\; -$   &       \\
       
       \hline
    $-1.611 + 3.244i$         &     $-1.611 - 3.244i$             & $-6.066$ &      $-2$&            $-0.660$ &           $-1.458$  &      $-1.898$&        $\;\;\;\; -$&       $\;\;\;\; -$&         $\;\;\;\; -$   &        \\
    \hline
    \hline
        $-1.644 + 3.119i$         & $-1.644 - 3.119i$               & $\;\;\;\; -$ & $\;\;\;\; -$            & $\;\;\;\; -$ &     $\;\;\;\; -$   &     $\;\;\;\; -$   &     $\;\;\;\; -$  &  $-2.797$   &     $\;\;\;\; -$   &     $\;\;\;\; -$    \\
    
      \hline
       $-1.624+3.139i$         & $-1.624 - 3.139i$               & $\;\;\;\; -$  & $\;\;\;\; -$            & $\;\;\;\; -$ &     $\;\;\;\; -$   &     $\;\;\;\; -$   &     $\;\;\;\; -$  &  $-3.131$   &     $-3.131$   &     $\;\;\;\; -$    \\
    
       \hline
    $-1.630+3.159i$         & $-1.630-3.159i$                   & $\;\;\;\; -$ & $\;\;\;\; -$            & $\;\;\;\; -$ &     $\;\;\;\; -$   &     $\;\;\;\; -$   &     $\;\;\;\; -$  &   $-3.465$   &     $-3.465$   &     $-3.465$    \\
 
    \hline
    \hline
   $-1.611+3.244i$      & $-1.611-3.244i$                   & $-6.066$  & $-2$    &     $\;\;\;\; -$   &     $\;\;\;\; -$   &     $\;\;\;\; -$  & $-4$ &     $\;\;\;\; -$ &     $\;\;\;\; -$   &     $\;\;\;\; -$   \\
   \hline
   $-1.611+3.244i$      & $-1.611-3.244i$                    & $-6.066$  & $-2$    &     $\;\;\;\; -$   &     $\;\;\;\; -$   &     $\;\;\;\; -$  & $-4$ &     $-2.780$ &     $\;\;\;\; -$   &     $\;\;\;\; -$   \\
    \hline
  \hline
  \end{tabular}
\caption{The UV fixed point- and corresponding eigen-values $\gamma_i$ for different truncations in the $\stuck$ sector of the action, for the functions $f(\psi)=\sum_{i = 0}^{N_f}\frac{1}{i!}\rho_{i}\psi^{i}$, $q(\psi) =  \sum_{i = 1}^{N_q}\frac{1}{i!}\sigma_{i}\psi^{i}$. Notice that when $f=0$, $q$ has to be at least of second order in $\psi$, as explained in the text. The Einstein--Hilbert sector, spanned by the couplings $G$ and $\Lambda$ is held fixed. The couplings $G$ and $\Lambda$ posses a non--trivial and attractive UV fixed point, for all the truncations considered, with a value that appears to be stable and very close to that of GR. Notice also that for all cases considered, including standard GR, the dimensionless product $\tilde{G}_{*}\tilde{\Lambda}_{*}$ retains the same value up to two decimal places. Furthermore, the couplings of the $\stuck$ sector become trivial in the UV (i.e flow to zero), and they all appear to be attractive (i.e relevant) as the truncation order is increased, as can be seen from the corresponding eigenvalues, which are calculated from the linearisation matrix (\ref{linear-matrix}). The attractivity properties of the $\stuck$ sector change after canonically normalising the dimensionless field $\psi$ to mass dimension one,
as discussed in the text. In that case, the associated eigenvalues are shifted as $\gamma_{\tilde{\rho}_n, \tilde{\sigma}_n} \to \gamma_{\tilde{\rho}_n, \tilde{\sigma}_n}+n$, and for all truncations studied it turns out that there is a point where the critical exponents turn from relevant to irrelevant, suggesting a finite dimensional critical manifold for the unimodular theory.}\label{maintable}
\end{table*}

\section{Discussion and conclusions} \label{sec:Conclusions} 
We studied the quantum dynamics of unimodular gravity, starting from a unimodular formulation of GR which is by construction invariant under diffeomorphism transformations. The action consisted, apart from the metric field, of five new field variables, namely a Lagrange multiplier $\lambda$ and a set of four $\stuck$ scalars $\phi^{\alpha}$. As expected, the theory is equivalent to GR at the classical level. 

By using the tool of the functional RG, and in particular evaluating an exact RG equation, the UV properties of the corresponding effective action were studied, under the assumptions described in section (\ref{sec:part2}). In particular, we allowed the Lagrange multiplier and $\stuck$ fields to couple with appropriate sources in the generating functional, therefore treating them on equal footing with the metric field, which is a conceptually different approach than that suggested in Ref. \cite{Padilla:2014yea}.  The $\stuck$ interactions expanded the theory space of the Einstein--Hilbert action through the corresponding couplings. What is more, our analysis included a cosmological constant, since not only there was no symmetry to prevent its inclusion, but the in principle off--shell nature of quantum fluctuations calculated would be expected to violate the on--shell unimodularity condition. The first main result of this work is the RG flow equation presented in equation (\ref{ERGE-rhs}), while the numerical results for the fixed points and associated eigenvalues can be found in Table \ref{maintable}.

Let us summarise the key outcomes of our analysis below:

\begin{itemize}
\item In the extreme UV (continuum) limit corresponding to the IR cut--off taken to infinity, i.e $k \to \infty$, and for the cases we considered in this work, the  $\stuck$ sector of the unimodular theory became trivial, i.e $\{\tilde{\rho}_{i}, \tilde{\sigma}_i \} \to 0$, while the Einstein--Hilbert sector, spanned by Newton's and cosmological constant respectively, appeared to be asymptotically safe with an attractive fixed point very close in value to that of standard GR, and the same turned out to be true for the associated eigenvalues.  The results appeared to be stable under increasing the order of the $\stuck$ interactions. 
The quantitative similarity between the two Einstein--Hilbert sectors, in combination with the vanishing of the $\stuck$ sector in the unimodular theory, suggests that at sufficiently high energies the effective action of the unimodular theory  acquires essentially a similar form with that of standard GR, and we can write
\be
\left. \Gamma^{\text{Unim.}} \right |_{k/k_{0} \gg 1}\simeq  \left. \Gamma^{\text{GR}} \right |_{k/k_{0} \gg 1}. \label{EffActionCompare}
\ee
In this sense, the two theories appear to share the same UV completion within the context of asymptotic safety. What is more, for the dimensionless product $G \times \Lambda$ evaluated on the fixed point, we found that
\begin{align}
\left.  \tilde{G}_{*}\tilde{\Lambda}_{*} \right|_{\text{Unimod.}} \simeq \left. \tilde{G}_{*}\tilde{\Lambda}_{*} \right|_{\text{GR}}.
\end{align} 
We should point out that the similarity between the two theories in the UV found here is not to be a priori expected; in fact, for the unimodular theory we considered, the unimodularity condition appears as an on--shell condition, while the quantum corrections calculated through the exact RG equation are in principle off--shell. What is more, the similarity between the two UV completions does not imply their equivalence at the quantum level; in the Wilsonian context we considered in this work, this can be seen by the fact that the RG flow will in principle generate non--trivial, novel compared to GR interactions in the $\stuck$ sector as the cut--off scale $k$ is lowered towards the IR. In this case, the relation (\ref{EffActionCompare}) will break down and the two effective actions will no longer share the same structure. This is an immediate result of allowing the Lagrange multiplier and $\stuck$ fields to couple directly to the sources in the path integral (\ref{PI}). We shall leave the phenomenological implications of this issue for future work. 
 
\item On the same time, we found that the predictivity of the unimodular theory seems to be endangered by the apparent increase of relevant couplings as the truncation order of the $\stuck$ sector is increased. The root of this potentially problematic behaviour appears to be the dimensionless nature of the field $\psi$ in the original unimodular action.
 In fact, and for all truncations we considered, it turns out that canonically normalising the field, which could be further motivated by the inclusion of a kinetic term for $\psi$ in the effective action, introduces a point in the truncation order where the critical exponents turn from relevant to irrelevant, indicating the existence of only a finite number of relevant couplings in the $\stuck$ sector, and on the same time ensuring that the theory is predictive.

\end{itemize}

\acknowledgements
The author is thankful to Tony Padilla for numerous enlightening discussions, useful suggestions and contributions at earlier stages of this work, as well as for providing an independent check of part of the calculations. Furthermore, he would like to thank Dario Benedetti,  J\"urgen Dietz, Astrid Eichhorn, Kai Groh, Daniel Litim, Tim Morris, Kostas Nikolakopoulos, Christoph Rahmede and Frank Saueressig for valuable discussions and correspondence. The author gratefully acknowledges financial support by an STFC consolidating grant from the University of Nottingham, as well as by the Fundação para a Ciência e Tecnologia Investigador FCT research grant SFRH/BPD/95204/2013. He further acknowledges project PEst-OE/FIS/UI2751/2014 (CAAUL).

\appendix

\section{Hessian entries on $S_4$} \label{hessian}

To calculate the expanded effective action up to second order we expand the field variables into a background and a longitudinal and transverse perturbation piece respectively as,
\begin{eqnarray}
&&g_{\mu \nu}(x)  =  \bar{g}_{\mu \nu}(x) + G_{0}^{1/2} \hat{h}_{\alpha \beta}(x) + \frac{1}{4}G_{0}^{1/2}\bar{g}_{\alpha \beta}h(x), \nonumber \\
&&\lambda(x) = \bar{\lambda}(x) + G_{0}^{-1/2} \delta \lambda(x), \no \\
&& \phi^{\alpha}(x) = \bar{\phi}^{\alpha}(x) + G_{0}^{1/2}\delta \hat{\phi}^{\alpha}(x)  +  G_0 \bar \nabla^{\alpha} \delta \phi(x).
\end{eqnarray}

Restricting the background to be that of a Euclidean sphere $S_4$, and using above decompositions, the explicit Hessian entries are given as follows
\begin{align}
 & \hspace{-1.5cm} \left[ \Gamma^{(2)}_{(\hat{h}\hat{h})}\right]^{\mu \nu}_{\rho \sigma}
 = Z_{\hat{h}\hat{h}} \left[ -\Box  - 2\Lambda   + \frac{2}{3}R -  Z_{G}^{-1} \left( \lambda V(\psi) + U(\psi) \right)  \right] P^{\mu \nu}_{\rho \sigma}, \no \\
 \vspace{0.1cm} \no \\
&\hspace{-1cm} \Gamma^{(2)}_{(hh)} = Z_{hh}\Big[  -\Box  -  2\Lambda   - Z_{G}^{-1}\left( \lambda V(\psi) + U(\psi) \right) \no \\
& - 2Z_{G}^{-1} \psi^2 \left( \lambda f'' +q'' \right)   \Big] , \no \\
 \vspace{0.1cm} \no \\
& \hspace{-1.5cm} \Gamma^{(2)}_{(\phi \phi)} =  -G_{0}^3 \Big[ \frac{1}{4}\left(\frac{\psi}{\epsilon^2} \right)\left( \lambda f' + q' \right)R (-\Box) \no \\
&   + \left( \frac{\psi^2}{\epsilon^2} \right) \left( \lambda f'' + q'' \right)(-\Box)(-\Box) \Big] , \no \\
 \vspace{0.1cm} \no \\
\left[ \Gamma^{(2)}_{(\hat{\phi} \hat{\phi})} \right]_{\mu \nu} &=   -\frac{G_{0}^2}{4}  \left(\frac{\psi}{\epsilon^2} \right) \left( \lambda  f' + q'  \right) \,  R \,  g_{\mu \nu} , \no \\
 \vspace{0.1cm} \no \\
\Gamma^{(2)}_{(h \lambda)} &=  -\frac{1}{2} V(\psi),\no\\
%
   \Gamma^{(2)}_{(\phi \lambda)} &= 
G_{0} \frac{\psi}{\epsilon}  f' (-\Box),\no  \\
  \Gamma^{(2)}_{( h \phi )}  &=-\frac{1}{2}G_{0}^{2} \frac{\psi^2}{\epsilon} \left(  \lambda f'' + q'' \right) (- \Box),
\label{Hessian-entries}
\end{align}
with the corresponding wave function renormalisations defined as
\be
Z_{\hat{h}\hat{h}} \equiv  \frac{1}{2}G_{0}Z_{G}, \; \; Z_{hh}\equiv  - \frac{1}{8}G_{0} Z_{G},
\ee
the projector $P^{\mu \nu}_{\rho \sigma}$ given by
\be
P^{\mu \nu}_{\rho \sigma} \equiv \delta^{\mu \nu}_{\rho \sigma} - \frac{1}{4} g^{\mu \nu} g_{\rho \sigma},
\ee
with $\delta^{\mu \nu}_{\rho \sigma} \equiv \frac{1}{2}(\delta^{\mu}_{\rho} \delta^{\mu}_{\sigma} + \delta^{\mu}_{\sigma} \delta^{\mu}_{\rho})$,
and $V(\psi) \equiv f(\psi) - \psi f'(\psi), \; \; U(\psi) \equiv q(\psi) - \psi q'(\psi)$. Also, notice that the value of the scalar $\psi$ on the background is
\be
\psi = \frac{\epsilon^4}{\sqrt{\bar{g}}}.
\ee
We also remind the reader that the Laplacian $\Box$ is constructed out of the background metric, $\Box \equiv \bar{g}^{\mu \nu} \bar{\nabla}_{\mu} \bar{\nabla}_{\nu}$, and that $\Lambda \equiv \Lambda(k)$, $G \equiv G(k)$ and $f \equiv f(\psi,k)$, $q(\psi) \equiv q(\psi,k)$ and $' \equiv d/d\psi$.

\section{$\stuck$ ghost} \label{ghost}
To calculate the ghost term from the St\"uckelberg field decomposition into transverse and longitudinal part, we first need to calculate the determinant associated with the decomposition. We start with the Gaussian integral over the St\"uckelberg fields $\phi^{\alpha}$ and then perform the decomposition in the exponential as,
\begin{align}
&\int \mathcal{D}\phi^{\alpha} e^{- \frac{1}{2} <\phi^{\alpha}, \phi^{\alpha}>} = J_{\phi} \int \mathcal{D}\hat{\phi}^{\alpha} \mathcal{D}\phi^{\alpha} e^{- \frac{1}{2} \left[ \hat{\phi}^{\alpha}\hat{\phi}_{\alpha} - \phi \Box \phi\right]} \nonumber \\
& \equiv  J_{\phi} \int \mathcal{D}\hat{\phi}^{\alpha}e^{- \frac{1}{2} \hat{\phi}^{\alpha}\hat{\phi}_{\alpha}} \int \mathcal{D}\phi^{\alpha} e^{- \frac{1}{2} \left(- \phi \Box \phi\right)},
\end{align}
with the measure $\int \mathcal{D}\phi^{\alpha}$ denoting summation over all field configurations. 
We then find for the determinant $J_{\phi}$ that
\be
J_{\phi} = \left( \text{det}(-\Box) \right)^{1/2}.
\ee
Introducing the scalar ghost and anti-ghosts $\eta$ and $\bar{\eta}$ respectively, we can represent the Jacobian $J_{\phi}$ according to the Fadeev--Popov procedure \cite{Peskin} as
\be
J_{\phi} =  \int  \mathcal{D}\eta  \mathcal{D} \bar{\eta} e^{- \frac{1}{2} \bar{\eta} (-\Box) \eta },
\ee
which implies that the associated ghost action is
\be
S_{\eta}^{\text{(ghost)}} \equiv - \frac{1}{2}\int \bar{\eta} (- \Box) \eta.
\ee
Notice that we have assumed that the wave function renormalisation for the ghost action does not run and that it is equal to one.

\section{Trace integrals} \label{trace}
The ERGE requires us to calculate traces of functions of the Lapacian operator, which we perform asymptotically making use of an appropriate heat kernel expansion. 

With the definitions $z \equiv -\Box$, the modified Laplacian $P_k(z) \equiv z + R_{k}(z)$, where $R_{k}(z)$ is the regulator profile function, and ${\bf U}$ an operator independent of $z$, but in principle depending on the curvature $R$ and the (background) fields of the action, the trace integrals can be asymptotically expanded in the small-scale regime where $\bar{R}/k^2, \bar{\lambda}/k^2 \ll 1$, with $ \bar{R} = \text{constant}$, $\bar{\lambda} = \text{constant}$, as
\begin{align}
 \text{Tr} & \left[ \frac{k\partial_{k} g(z) }{P_{k}(z) + {\bf U}} \right]  =   \frac{1}{(4\pi)^{2}} \sum_{i}^{\infty} \sum_{l = 0}^{\infty} Q_{2-i}\left(\frac{k\partial_k g(z) }{P_{k}^{l+1}(z)} \right) \times \no \\
& \times (-1)^{l}  \int d^{4}x\sqrt{g} \, \text{tr}({\bf U})^{l} \, b_{2i}(z), \label{HK-expansion}
\end{align}
with the function $g(z)$ collectively denoting the different cases $g(z) \equiv R_{k}$ and $g(z) \equiv k\partial_k R_k$, with the regulator function chosen to be the ''optimised" (Litim's) one, $R_{k}(z) = (k^2 - z)\Theta(k^2 - z)$ \cite{Litim_Opt-CO1}, with $\Theta$ the Heaviside step function. What is more, the symbol ''$\text{Tr}$" stands for the trace over space-time indices and momenta. The functionals $Q_{2-i}$ in (\ref{HK-expansion}) are defined as
\be
Q_{2-i}\left(\frac{k\partial_k g(z)}{P_{k}^{l+1}(z)} \right) \equiv Q_{2-i}(z) = \int_{0}^{\infty}ds e^{-zs}\tilde{g}(z).
\ee 
$\tilde{g}(z)$ denotes the anti-Laplace transformed $g(z)$, and the quantities $b_{2i}(z)$ are closely related to the heat kernel coefficients of the Laplacian $z \equiv - \Box$. For more explicit details we refer to Refs \cite{Reuter:1996cp,Dou:1997fg, Reuter:2012id,Reuter:2007rv,Lauscher:2001ya,Cod-Perc-Rahme1}.

\bibliography{UM.bib,ASreferences.bib,AS_inflation.bib}

\end{document}